\documentclass[twocolumn,showpacs,prl,superscriptaddress,preprintnumbers]{revtex4}
\usepackage{graphicx}

\begin{document}
\title{Generalized spin polarizabilities of the nucleon in Heavy
Baryon Chiral Perturbation Theory at order ${\mathcal O}(p^4)$}
\author{Chung Wen Kao} 
\affiliation{Department of Physics and Astronomy, University of
Manchester, Manchester, M13 9PL UK}
\author{Marc Vanderhaeghen}
\affiliation{Institut f\"{u}r Kernphysik, Johannes Gutenberg-Universit\"{a}t, D-55099 Mainz, Germany}
\date{\today}
\begin{abstract}
We present the first heavy baryon chiral perturbation theory
calculation at order ${\cal O}(p^4)$ for the spin-dependent amplitudes 
for virtual Compton scattering off the nucleon, 
and extract the ${\cal O}(p^4)$ results for the 
generalized spin polarizabilities of the nucleon. 
\end{abstract}
\maketitle
Over the past years, the virtual Compton scattering (VCS) process on
the nucleon, accessed through the $e p \to e p \gamma$ reaction, has
become a powerful new tool to study the internal structure of the
nucleon both at low and high energies (see \cite{GV98} for a review). 
Being an electromagnetic process, it provides us with a well understood probe,
as the electron interacts with the nucleon, to a very good precision, 
through the exchange of one (virtual) photon. Furthermore, in contrast
to the real Compton scattering (RCS) process, VCS allows us to vary
the virtuality of the initial photon, which plays the role of the
resolution with which one ``sees'' the constituents within the
nucleon. At very high virtuality and energy, the photon interacts with
a single quark in the nucleon, and one accesses Compton scattering at
the quark level. The VCS process on the nucleon tells us then how this 
quark is embedded in the nucleon, which is parametrized through
so-called generalized parton distributions. 
At low virtuality and energy, the outgoing real photon plays the role
of an applied quasi-static electromagnetic field, and the VCS process
measures the response of the nucleon to this applied field, which can
be parametrized in nucleon structure quantities, termed 
generalized polarizabilities (GPs) \cite{GLT}. 
In this case, the virtuality of the initial photon can be dialed so as
to map out the spatial distribution of the electric polarization, the
magnetization, or the spin densities of the nucleon~\cite{spatial}. 
In particular, this allows us to study, at different length scales,
the interplay of the role of the pion cloud and quark core 
contributions to the nucleon GPs. 
Because the chiral dynamics plays a dominant role in this regime, 
at low virtuality,  
Heavy Baryon Chiral Perturbation Theory (HBChPT) provides a natural
way to study the GPs of the nucleon.
\newline
\indent
On the experimental side, the measurement of the VCS process
became only possible in recent years, with the advent of high
performance electron accelerator facilities, as these experiments
require accurate measurements of small cross sections. At low photon
energy and virtuality, 
first unpolarized VCS observables have been measured  
at the MAMI accelerator \cite{Roc00} at a virtuality $Q^2$ = 0.33 GeV$^2$, 
and recently at JLab \cite{Fon02} at higher virtualities, 
$1 < Q^2 < 2~$GeV$^2$, as well as at MIT-Bates \cite{Mis97}.
All those experiments measure two combinations of GPs. 
\newline
\indent
The nucleon structure functions extracted from the 
first unpolarized VCS experiment \cite{Roc00}, at a moderately large
virtuality of $Q^2$ = 0.33 GeV$^2$, are in surprisingly good
agreement with the ${\cal O}(p^3)$ HBChPT predictions for the GPs,
performed in Refs.~\cite{HHKS,HHKD}.
In comparison with other model approaches, the good agreement between the
unpolarized VCS response functions of \cite{Roc00} and the ${\cal O}(p^3)$
HBChPT result, is for an important part due to the large spin 
GPs of the nucleon at this order, which also contribute to the
unpolarized observables. This has triggered a further 
experimental program~\cite{dHo01} to directly access and separate the
spin GPs of the nucleon by measuring VCS double polarization
observables~\cite{doublepol}. 
To sort out the importance of the spin GPs of the nucleon, we
calculate in this letter the spin-dependent VCS amplitudes in HBChPT at order
${\cal O}(p^4)$. At this order, no unknown low-energy constants enter
the calculations so that one can extract 
the ${\cal O}(p^4)$ results for the 
spin GPs of the nucleon, without free parameters.    
\newline
\indent
We begin by specifying our notation for the VCS process:
\begin{equation}
\gamma^{*}(\epsilon_{1},q)+N(p_{1})\rightarrow\gamma(\epsilon_{2},q')+N(p_{2}).
\end{equation} 
In calculating the VCS amplitudes, we work in the c.m. frame and 
choose the Coulomb gauge.
In the VCS process, the initial spacelike photon is characterized 
by its four momentum $q=(\omega,\vec{q})$, virtuality $Q^2 \equiv - q^2$, 
and polarization vector $\epsilon_{1}=(0,\vec{\epsilon}_{1})$.
The outgoing real photon has four momentum 
$q'=(\omega'=|\vec{q'}|,\vec{q'})$ and 
polarization vector $\epsilon_{2}=(0,\vec{\epsilon}_{2})$.
We define $\bar{q}\equiv|\vec{q}|$, and denote $\theta$ as the
scattering angle between virtual and real photons, i.e., 
$\cos\theta=\hat{q}\cdot\hat{q}^{\, '}$.
The polarization vector $\vec{\epsilon}_{1}$ of the virtual photon 
can be decomposed into a longitudinal component 
$\vec{\epsilon}_{1L}=(\vec{\epsilon}_{1}\cdot\hat{q})\hat{q}$ and 
a transverse component $\vec{\epsilon}_{1T}$. 
For further use, we introduce the virtual photon energy in the limit 
$\omega' = 0$~:
\begin{eqnarray}
\omega_{0}\equiv \omega(\omega'=0,\bar{q})&=&
M_{N}-\sqrt{M_{N}^{2}+\bar{q}^{2}},
\nonumber\\
&=& -{\bar{q}^2} / (2 M_{N}) + {\mathcal O}(1/M_{N}^{2}),
\label{eq:omegao}
\end{eqnarray}
with $M_N$ the nucleon mass, and where the 
last line in Eq.~(\ref{eq:omegao}) indicates the heavy baryon expansion.
\newline
\indent
The VCS amplitude ${\cal M}_{VCS}$ 
can be expressed in terms of twelve structure functions as \cite{HHKS}~: 
\begin{eqnarray}
&&{\cal M}_{VCS} \,=\, i e^2 \left\{
(\vec{\epsilon}_{2}^{\, *}\cdot \vec{\epsilon}_{1T})A_{1}
+(\vec{\epsilon}_{2}^{\, *}\cdot\hat{q})
(\vec{\epsilon}_{1T}\cdot\hat{q'})A_{2} \right. \nonumber\\
&&+\, i\vec{\sigma}\cdot(\vec{\epsilon}_{2}^{\, *}\times\vec{\epsilon}_{1T})A_{3}
+i\vec{\sigma}\cdot(\hat{q'}\times\hat{q})
(\vec{\epsilon}_{2}^{\, *}\cdot \vec{\epsilon}_{1T})A_{4} \nonumber \\
&&+\, i\vec{\sigma}\cdot(\vec{\epsilon}_{2}^{\, *}\times\hat{q})(\vec{\epsilon}_{1T}\cdot\hat{q'})A_{5}
+i\vec{\sigma}\cdot(\vec{\epsilon}_{2}^{\,*}
\times\hat{q'})(\vec{\epsilon}_{1T}\cdot\hat{q'})A_{6} 
\nonumber\\
&&-\, i\vec{\sigma}\cdot(\vec{\epsilon}_{1T}\times\hat{q'})(\vec{\epsilon}_{2}^{\,*}\cdot\hat{q})A_{7} 
-i\vec{\sigma}\cdot(\vec{\epsilon}_{1T}\times\hat{q})(\vec{\epsilon}_{2}^{\,*}\cdot\hat{q})A_{8} \nonumber \\
&&+\, (\vec{\epsilon}_{1L}\cdot \hat{q}) \left[
(\vec{\epsilon}_{2}^{\,*}\cdot\hat{q}) A_{9}
+i\vec{\sigma}\cdot(\hat{q'}\times\hat{q})
(\vec{\epsilon}_{2}^{\,*}\cdot\hat{q}) A_{10} \right. \nonumber\\
&&\left. \left. \hspace{1.4cm} 
+ \, i\vec{\sigma}\cdot(\vec{\epsilon}_{2}^{\,*}\times\hat{q}) A_{11}
+i\vec{\sigma}\cdot(\vec{\epsilon}_{2}^{\,*}\times\hat{q'}) A_{12}
\right] \right\} . 
\label{eq:vcsampl}
\end{eqnarray}
To extract the GPs, we first calculate the complete fourth order VCS 
amplitudes $A_i$ in HBChPT, and subsequently separate the amplitudes
in a Born part $A_{i}^{Born}$, and a non-Born part $\bar{A}_{i}$ as~:
\begin{eqnarray}
A_{i}(\omega',\bar{q},\theta)=A_{i}^{Born}(\omega',\bar{q},\theta)
+\bar{A}_{i}(\omega',\bar{q},\theta),
\end{eqnarray}
In the Born process, the virtual photon is absorbed on
a nucleon and the intermediate state remains a nucleon. 
We calculate them, following the definition of Ref.~\cite{GLT}, 
from direct and crossed Born diagrams with the
electromagnetic vertex given by~:
$$\Gamma^{\mu}(q^2)=F_{1}(q^2)\gamma^{\mu}+F_{2}(q^2)i\sigma^{\mu\nu}\frac{q_{\nu}}{2M_{N}},$$
where $F_{1}(F_{2})$ are the nucleon Dirac (Pauli) form factors respectively.
The non-Born terms contain the information 
on the internal structure of the nucleon. 
In the framework of HBChPT, the non-Born amplitudes start at order 
${\cal O}(p^3)$, where $p$ ($p = m_\pi, \bar q, \omega'$) 
is a momentum scale of the problem,  
which is small compared with the nucleon mass $M_N$, 
such that a systematic expansion can be made in powers of $p / M_N$.   
In particular, we are interested in this work 
in the nucleon response to an 
applied quasi-static electromagnetic dipole field. 
This is accessed by performing 
a low energy expansion (LEX), in the outgoing photon energy $\omega'$,  
of the non-Born VCS amplitudes, and by selecting the  
term linear in $\omega'$, as~:
\begin{equation}
\left[\frac{\partial \bar{A}_{i}}{\partial \omega'}(\omega',\bar{q},\theta)\right]_{\omega'=0}
={\cal S}_{i}(\bar{q})+{\cal P}_{i}(\bar{q})\cos\theta .
\label{eq:linterm}
\end{equation}
In Eq.~(\ref{eq:linterm}), 
the nucleon structure quantities ${\cal S}_{i}$ and 
${\cal P}_{i}$ can be expressed in terms of the GPs of the
nucleon, as introduced in~\cite{GLT}, 
which are functions of $\bar q$ and which are denoted by
$P^{(\rho' \, L', \rho \,L)S}(\bar q)$.
In this notation, $\rho$ ($\rho'$) refers to the
electric ($E$), magnetic ($M$) or longitudinal ($L$) nature of the initial
(final) photon, $L$ ($L'$) represents the angular momentum of the
initial (final) photon, and $S$ differentiates between the
spin-flip ($S=1$) and non spin-flip ($S=0$)
character of the transition at the nucleon side.
Restricting oneself to a dipole transition for the final photon
(i.e. $L'$ = 1), angular momentum and parity conservation lead to
3 scalar and 7 spin GPs \cite{GLT}.
The 7 spin GPs, which are the subject of investigation in this letter, 
are obtained as \cite{Mainz1}~:
\begin{eqnarray}
P^{(M1,L2)1}&=&
\frac{-2\sqrt{2}}{3\sqrt{3}} \frac{1}{\bar{q} \, \omega_0}
\sqrt{{{M_N} \over E_N}} 
{\cal S}_{10},
\nonumber\\
P^{(L1,L1)1} &=& \frac{-2}{3} \frac{1}{\omega_0} \sqrt{{{M_N} \over E_N}} 
{\cal S}_{11}, 
\nonumber \\
P^{(M1,L0)1} &=& \frac{2}{\sqrt{3}} \frac{\bar{q}}{\omega_0} 
\sqrt{{{M_N} \over E_N}} 
\left[{\cal S}_{12} -\frac{2}{3} {\cal S}_{10}\right], 
\nonumber \\
P^{(L1,M2)1}&=&-\frac{\sqrt{2}}{3 \bar{q}^2} \sqrt{{{M_N} \over E_N}} 
{\cal S}_{8},
\nonumber \\
P^{(M1,M1)1}&=&\frac{2}{3 \bar{q}} \sqrt{{{M_N} \over E_N}} 
\left[{\cal S}_{7} -{\cal S}_{4}\right], 
\nonumber \\
\hat{P}^{(M1,2)1}&=&-\frac{4}{3\sqrt{10} \bar{q}^3} \sqrt{{{M_N} \over E_N}} 
\left[{\cal S}_{4} + {\cal S}_{7}-{\cal S}_{10}\right],
\nonumber \\
\hat{P}^{(L1,1)1}&=&
-\frac{2\sqrt{2}}{3\sqrt{3} \bar{q}^2} \sqrt{{{M_N} \over E_N}} 
\left[{\cal S}_{3}+\frac{1}{2}{\cal S}_{8}-{\cal S}_{11}\right] ,
\label{gp}
\end{eqnarray}
where $E_N = \sqrt{M_{N}^{2}+\bar{q}^{2}}$, and 
where the GPs denoted by $\hat P$ correspond with 
mixed electric and longitudinal
multipoles, as introduced in~\cite{GLT}.
We will call a GP as ${\cal O}(p^n)$ in HBChPT if it is extracted
from the ${\cal O}(p^n)$ VCS amplitudes using Eq.~(\ref{gp}).
As the linear term in the LEX of the 9 spin-dependent VCS
amplitudes of Eq.~(\ref{eq:vcsampl}) depends upon only 7
independent GPs, there is a redundancy among the coefficients of
Eq.~(\ref{eq:linterm}). In particular, one has the following  
constraints between the ${\cal S}_{i}$ and ${\cal P}_{i}$~:
\begin{eqnarray}
{\cal P}_{3}&=&-{\cal S}_{7},\hspace{0.5cm}
{\cal P}_{11}=-{\cal S}_{10}, \hspace{0.5cm} 
{\cal S}_{4}=-{\cal S}_{5}, \nonumber \\
{\cal S}_{6}&=&{\cal P}_{4}={\cal P}_{5}={\cal P}_{6}={\cal P}_{7}={\cal P}_{8}
={\cal P}_{10}={\cal P}_{12}=0, 
\label{con1}
\end{eqnarray}
which provide consistency checks for the HBChPT results of the
VCS amplitudes at a given order, and have been verified up to fourth
order here. Furthermore, it has been shown \cite{Mainz1} that
nucleon crossing combined with charge conjugation symmetry of the VCS
amplitudes provides 3 additional constraints among the 7 spin GPs~:
\begin{eqnarray}
{\cal S}_{4}=0, 
\hspace{0.75cm}
{\cal S}_{3}=\frac{\bar{q}}{\omega_0} {\cal S}_{7}, 
\hspace{0.75cm}
{\cal S}_{10}-{\cal S}_{12}=\frac{\bar{q}}{\omega_0} {\cal S}_{11},
\label{con2}
\end{eqnarray}
leaving 4 independent spin GPs. 
By expanding $\omega_{0}$ as in~(\ref{eq:omegao}), 
it is obvious that the relations (\ref{con2}) 
connect quantities of different order in the heavy baryon
expansion. For the first non-trivial check of these relations in 
HBChPT, one has to calculate the VCS amplitudes to ${\cal O}(p^4)$. 
This is performed for the first time for the spin-dependent VCS
amplitudes in this letter, and we verified that the VCS amplitudes in 
HBChPT at ${\cal O}(p^4)$ satisfy the relations~(\ref{con2}). 
\newline
\indent
Furthermore, in the limit $\bar{q}\rightarrow 0$, the GPs are 
related with the Ragusa spin polarizabilities \cite{Ragusa} of RCS, 
which have been calculated before in HBChPT at order ${\cal O}(p^4)$ 
\cite{Kum00,JKO,Gel00}. We verified that 
our ${\cal O}(p^4)$ results reduce at the real photon point to those
of Refs.~\cite{Kum00,JKO}.
\newline
\indent
By calculating the VCS amplitudes in HBChPT at ${\cal O}(p^3)$, 
one extracts from Eq.~(\ref{gp}) the following expressions for the ${\cal O}(p^3)$ GPs
(see Ref.~\cite{HHKD})~:

\begin{small}
\begin{eqnarray}
&&\hspace{-0.5cm} P^{(M1,L2)1}(\bar{q}) = 
P^{(M1,L0)1}(\bar{q}) \,=\, \;\;
\hat{P}^{(M1,2)1}(\bar{q}) \,=\,0, 
\nonumber \\
&&\hspace{-0.5cm} P^{(M1,M1)1}(\bar{q}) =
P^{(L1,L1)1}(\bar{q}) \,=\, 0,
\nonumber \\
&&\hspace{-0.5cm} P^{(L1,M2)1}(\bar{q})=
\frac{-g_{A}^2}{24\sqrt{2}\pi^{2}F_{\pi}^{2}\bar{q}^{2}}
\left[1-\frac{4}{w\sqrt{w^2+4}}\sinh^{-1}[\frac{w}{2}]\right], \nonumber \\
&&\hspace{-0.5cm} \hat{P}^{(L1,1)1}(\bar{q})=
\frac{g_{A}^2}{24\sqrt{6}\pi^{2}F_{\pi}^{2}\bar{q}^{2}}
\left[3-\frac{4w^2+12}{w\sqrt{w^2+4}}\sinh^{-1}[\frac{w}{2}]\right],  
\label{result1}
\end{eqnarray}
\end{small}

\noindent
where $w \equiv \bar{q} / m_{\pi}$, with $m_\pi$ the pion mass.
Furthermore, throughout this paper we use the values~:
$g_A = 1.267$, $F_\pi = 0.0924$~GeV, and $m_\pi = 0.14$~GeV.
\newline
\indent
Our calculation of the VCS amplitudes in HBChPT at ${\cal O}(p^4)$, 
yields the ${\cal O}(p^4)$ GPs, from Eq.~(\ref{gp}), as~:

\begin{small}
\begin{eqnarray}
&&\hspace{-0.45cm} P^{(M1,L2)1}(\bar{q})=\frac{-g_{A}^{2}}{12\sqrt{6}\pi^2 F_{\pi}^{2}\bar{q}^{2}}
\left[1-\frac{4}{w\sqrt{w^2+4}}\sinh^{-1}[\frac{w}{2}]\right] \nonumber \\
&&\hspace{-0.45cm} P^{(M1,L0)1}(\bar{q})=\frac{g_{A}^{2}}{12\sqrt{3}\pi^2 F_{\pi}^{2}}
\left[2-\frac{3w^2+8}{w\sqrt{w^2+4}}\sinh^{-1}[\frac{w}{2}]\right], 
\nonumber \\
&&\hspace{-0.45cm} P^{(M1,M1)1}(\bar{q})=\frac{g_{A}^{2}}{24\pi^2 F_{\pi}^{2}M_{N}}
\left[1-\frac{w^2+4}{w\sqrt{w^2+4}}\sinh^{-1}[\frac{w}{2}]\right],\nonumber \\ 
&&\hspace{-0.45cm} \hat{P}^{(M1,2)1}(\bar{q})=
\frac{-g_{A}^{2}}{24\sqrt{10}\pi^2 F_{\pi}^{2}M_{N}\bar{q}^{2}}
\left[3-\frac{2w^2+12}{w\sqrt{w^2+4}}\sinh^{-1}[\frac{w}{2}]\right], 
\nonumber \\
&&\hspace{-0.45cm} P^{(L1,L1)1}(\bar{q})=0, \nonumber \\
&&\hspace{-0.45cm} P^{(L1,M2)1}(\bar{q})=
\frac{g_{A}^{2}}{96\sqrt{2}\pi F_{\pi}^{2} \bar{q}^2} 
\cdot {{\bar q} \over {M_N}} \nonumber \\
&&\hspace{1.5cm} \times \left[\frac{1}{2w}+\frac{2w^2+4}{w(w^2+4)}
+(\frac{5}{4}-\frac{3}{w^2})\tan^{-1}[\frac{w}{2}] \right. \nonumber\\
&&\hspace{1.5cm} \left. +\tau_{3}
\left(\frac{1}{2w}+(\frac{1}{4}-\frac{1}{w^2})\tan^{-1}[\frac{w}{2}] \right)
\right], \nonumber \\
&&\hspace{-0.45cm} \hat{P}^{(L1,1)1}(\bar{q})=
\frac{-g_{A}^{2}}{96\sqrt{6}\pi F_{\pi}^{2} \bar{q}^2} 
\cdot {{\bar q} \over {M_N}} \nonumber \\
&& \hspace{1.5cm} \times \left[\frac{11}{2w}-\frac{2w^2+4}{w(w^2+4)}
-(\frac{25}{4}+\frac{9}{w^2})\tan^{-1}[\frac{w}{2}] \right. \nonumber\\
&&\hspace{1.5cm} \left.
+\tau_{3}\left(\frac{3}{2w}-(\frac{5}{4}+\frac{3}{w^2})\tan^{-1}[\frac{w}{2}]\right)\right] .
\label{eq:nlo}
\end{eqnarray}
\end{small}

We can get more predictions by use of the crossing relations~(\ref{con2}), 
which hold in general in a relativistic quantum
field theory, and which we verified here in HBChPT at ${\cal O}(p^4)$. 
By plugging in the fourth order amplitudes in (\ref{con2}), we can
extract fifth order predictions from 
${\cal S}_{3}^{(4)}= -2M_{N} / {\bar{q}} \, {\cal S}_{7}^{(5)},$ 
and ${\cal S}_{10}^{(4)}-{\cal S}_{12}^{(4)}=$
$-2M_{N} / {\bar{q}} \, {\cal S}_{11}^{(5)}$. 
In this way, these relations allow us to extract two
${\cal O}(p^5)$ spin GPs: 

\begin{small}
\begin{eqnarray}
P^{(M1,M1)1}(\bar{q})&=&\frac{-g_{A}^{2}\bar{q}}
{192\pi F_{\pi}^{2}M_{N}^{2}}
\left[\frac{3}{w}-(\frac{5}{2}+\frac{6}{w^2})\tan^{-1}[\frac{w}{2}]
\right. \nonumber\\
&+&\left.\tau_{3}\left(\frac{1}{w}-(\frac{1}{2}+\frac{2}{w^2})\tan^{-1}
[\frac{w}{2}]\right)\right], 
\label{eq:nnlo} \\
P^{(L1,L1)1}(\bar{q})&=&\frac{g_{A}^{2}}{48\pi^2
F_{\pi}^{2}M_{N}}\left[-1+\frac{2w^2+4}{w\sqrt{w^2+4}}\sinh^{-1}[\frac{w}{2}]\right]. \nonumber
\end{eqnarray}
\end{small}

\noindent
The direct verification of these two predictions 
would imply a two-loop calculation, which has not been done so far. 
Furthermore, it is interesting to mention that the ${\mathcal O}(p^3)$ 
calculation~\cite{HHKD} was able to obtain some of these ${\cal O}(p^4)$ and one ${\cal O}(p^5)$ 
results, assuming the crossing relations~(\ref{con2}), 
in particular the ${\cal O}(p^4)$ results for 
$P^{(M1,M1)1}$ and $\hat P^{(M1,2)1}$, 
as well as the ${\cal O}(p^5)$ result for $P^{(L1,L1)1}$. 
Also, the ${\cal O}(p^4)$ results for $P^{(M1,L2)1}$ and $P^{(M1,L0)1}$ 
were obtained in~\cite{HHKS,HHKD}, by performing the
LEX of the amplitudes $\bar A_i$ to second order in $\omega'$, and, by
isolating two terms in $\omega'^2$ which depend on
those spin GPs. Besides confirming these previous results, our genuinely
new predictions at ${\cal O}(p^4)$ are for $P^{(L1,M2)1}$ and $\hat P^{(L1,1)1}$,
and at ${\cal O}(p^5)$ for $P^{(M1,M1)1}$.
It is also worth noting that the 
relations~(\ref{con2}) allow to classify the GPs in two groups~:
\begin{eqnarray}
{\mathrm{G}} 1 
&=&\{P^{(M1,L2)1}, P^{(M1,L0)1}, P^{(M1,M1)1}, 
\hat{P}^{(M1,2)1} \}, \nonumber\\
{\mathrm{G}} 2 
&=&\{P^{(L1,L1)1}, {P}^{(L1,M2)1}, \hat{P}^{(L1,1)1} \}. 
\nonumber
\end{eqnarray}
It is interesting to observe that their analytical 
forms alternate from one order to the next. At ${\cal O}(p^3)$ one has~:
\begin{eqnarray}  
{\mathrm{G}} 1 :
1/\pi \cdot \tan^{-1}[\bar{q} \,/\, 2m_{\pi} ];
\hspace{0.25cm} 
{\mathrm{G}} 2 :
1/\pi^2 \cdot \sinh^{-1}[\bar{q} \,/\, 2m_{\pi} ].
\nonumber
\end{eqnarray}
At ${\cal O}(p^4)$, one has~:
\begin{eqnarray}  
{\mathrm{G}} 1 :
1/\pi^2 \cdot \sinh^{-1}[\bar{q} \,/\, 2m_{\pi}];
\hspace{0.25cm} {\mathrm{G}} 2 :
1/\pi \cdot \tan^{-1}[\bar{q} \,/\, 2m_{\pi}] .
\nonumber
\end{eqnarray}  
At ${\cal O}(p^5)$ one again has the same analytical structure as at ${\cal O}(p^3)$, etc.
The alternating analytical forms 
teach us how the crossing relations~(\ref{con2}), 
connecting the GPs between two different orders in
HBChPT, work. 
\begin{figure}[h]
\vspace{-0.2cm}
\includegraphics[width=7.cm]{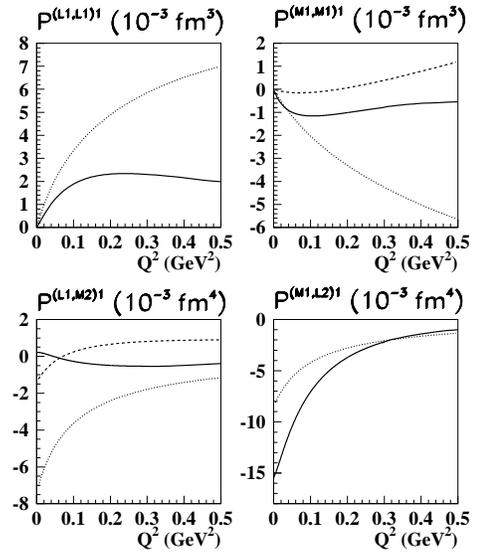}
\vspace{-0.4cm}
\caption[]{Results for the spin GPs :  
LO HBChPT results~\cite{HHKS,HHKD} (dotted curves);
NLO HBChPT results (dashed curves); 
dispersive evaluations of Ref.~\cite{Pas00,Pas01} (solid curves).
}
\label{fig:polarizab_comp}
\end{figure}
\newline
\indent
In the following, we denote the leading order non-vanishing results for the GPs
in HBChPT as LO HBChPT, and refer to the results including the
first chiral corrections as NLO HBChPT.   
In Fig.~\ref{fig:polarizab_comp} we show the 4 independent spin
GPs, and compare the LO HBChPT results of~\cite{HHKS,HHKD}, 
with the NLO results from the present work.
Furthermore, we also show a comparsion with a dispersion relation (DR)
evaluation~\cite{Pas00,Pas01}. 
The GPs $P^{\left(M 1, L 0\right)1}$, 
$P^{\left(M 1, L 2\right)1}$ and $P^{\left(L 1, L 1\right)1}$ 
are expected to be promising observables to study the
effects of the pion cloud surrounding the nucleon. 
Indeed, in a non-relativistic constituent quark model (CQM)
calculation~\cite{PSD01}, these GPs are vanishingly
small. This is because in  
a non-relativistic CQM, the only response to an applied 
static magnetic field is the alignment of the quark spins, whereas the
electric charge (or quadrupole) density remain unchanged. 
The HBChPT calculations in Fig.~\ref{fig:polarizab_comp} indeed show 
large contribution of pionic effects for these GPs. 
For $P^{\left(L 1, L 1\right)1}$ this result is confirmed by our 
${\mathcal O}(p^4)$ calculation. 
For the GP $P^{\left(M 1, M 1\right)1}$, one sees 
that the NLO result yields a large reduction compared to
the LO one, and calls the convergence of the
HBChPT result for this observable into question. 
Such a reduction has also been noticed in the 
linear $\sigma$-model calculation~\cite{LSM}, 
which takes account of part of the higher order
terms of a consistent chiral expansion, resulting in general in smaller values
for $P^{\left(L 1, L 1\right)1}$ and $P^{\left(M 1, M 1\right)1}$ 
compared with the LO calculations in HBChPT. 
For the GP $P^{\left(L 1, M 2\right)1}$,
we also notice that the NLO HBChPT result 
yields a relatively large correction. 
It is of course no surprise that the NLO correction are very large since the
corresponding calculations at the real photon point 
have already shown it \cite{Kum00,JKO}.
However it is worth noting that the 
NLO results show a smoother $Q^2$ dependence, and tend to lie closer
to the phenomenological DR estimates.

\begin{figure}[h]
\vspace{-.6cm}
\includegraphics[width=7.cm]{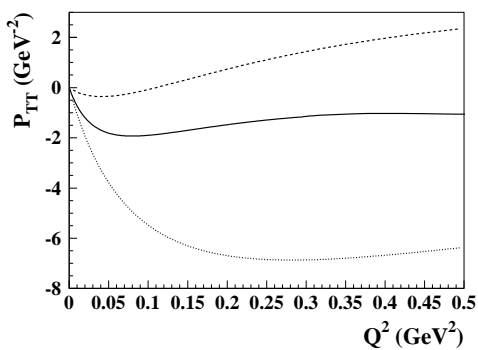}
\vspace{-0.3cm}
\caption[]{
Results for the VCS structure function $P_{TT}$. 
Dotted curve : LO HBChPT~\cite{HHKS,HHKD};
dashed curve : NLO HBChPT; 
solid curve : dispersive evaluation~\cite{Pas00,Pas01}.
}
\label{fig:ptt}
\end{figure}
\indent
The unpolarized VCS observables at low energy 
can be expressed in terms of 3 structure functions, one of which,
denoted by $P_{TT}$, involves only spin GPs \cite{GLT}~:
\begin{eqnarray}
P_{TT} \,=\, -3 \, G_{M} \,\frac{\bar{q}^{2}}{\omega_{0}}
\left(P^{(M1,M1)1}-\sqrt{2}\omega_{0}P^{(L1,M2)1}\right),
\end{eqnarray}
\noindent
with $G_M = F_1 + F_2$. 
We show the result for $P_{TT}$ in HBChPT in Fig.~\ref{fig:ptt}, and see that
it receives a very sizeable change at NLO. The measurement of 
$P_{TT}$ is planned in the near future \cite{dHo01} and 
will provide an interesting check on the spin densities of the nucleon.
\newline
\indent
In this work, we provided the first HBChPT calculation at ${\cal O}(p^4)$ 
for the spin-dependent VCS amplitudes. We extracted analytical
formulas for the ${\mathcal O}(p^4)$ spin GPs of the nucleon, 
which involve no free parameter at this order. 
Our results show sizeable chiral corrections for the spin GPs, which
are important for the interpretation of existing and forthcoming
experiments. 
\newline
\indent
This work was supported by the Deutsche Forschungsgemeinschaft (SFB443).
The authors also like to thank for the support and hospitality
of the ECT* (Trento), where this work was discussed.
Furthermore, the authors like to thank T. Hemmert and B. Pasquini 
for discussions.

\end{document}